\documentclass{Interspeech2024}
\usepackage{multirow}
\usepackage{algorithmic}
\usepackage{algorithm}

\interspeechcameraready

\title{Vision Transformer Segmentation for Visual Bird Sound Denoising}

\name[affiliation={1}]{Sahil}{Kumar}
\name[affiliation={2}]{Jialu}{Li}
\name[affiliation={1}]{Youshan}{Zhang}

\address{
  $^1$\small Department of Artificial Intelligence and Computer Science,  Yeshiva University, New York, NY, USA\\
  $^2$\small School of Public Policy, Cornell University, Ithaca, NY, USA. }
\email{skumar4@mail.yu.edu, jl4284@cornell.edu,youshan.zhang@yu.edu}


\begin{document}

\maketitle

\begin{abstract}
    

Audio denoising, especially in the context of bird sounds, remains a challenging task due to persistent residual noise. Traditional and deep learning methods often struggle with artificial or low-frequency noise. In this work, we propose ViTVS, a novel approach that leverages the power of the vision transformer (ViT) architecture. ViTVS adeptly combines segmentation techniques to disentangle clean audio from complex signal mixtures. Our key contributions encompass the development of ViTVS, introducing comprehensive, long-range, and multi-scale representations. These contributions directly tackle the limitations inherent in conventional approaches. Extensive experiments demonstrate that ViTVS outperforms state-of-the-art methods, positioning it as a benchmark solution for real-world bird sound denoising applications. Source code is available at: \url{https://github.com/aiai-4/ViVTS}.

\end{abstract}


\section{Introduction}

Audio denoising remains a persistent challenge in modern applications such as teleconferences, social media speech-to-text, and hearing aids~\cite{ephraim1985speech}. In the context of bird sound processing, the intricate and varied nature of avian vocalizations presents unique challenges. When addressing these challenges, traditional and deep learning methods are always constrained by issues like artificial noise and low-frequency noise. This has fueled the exploration of innovative solutions tailored for effective bird sound denoising~\cite{krishnamurthy2009babble}.

Audio denoising methods encompass a spectrum of representations including, time-domain, frequency-domain, and time-frequency domain approaches. Traditional methods, such as the Gaussian mixture model~\cite{Weiguang}, rely on statistical estimation to extract clean audio from noisy signals. Time-domain algorithms utilize techniques such as Short-Time Fourier Transform (STFT)~\cite{zhang2023birdsoundsdenoising} and Inverse Short-Time Fourier Transform (ISTFT)~\cite{wang2021tstnn}. Additionally, denoising performance is enhanced by applying Wiener filters and LSA estimators~\cite{wang2021denoising, lin2022multimodal}. Deep learning models, such as full convolutional neural networks (FCN), IBM-based DNNs, and Noise2Noise methods, have demonstrated superiority in audio denoising~\cite{gul2023survey}. These models exhibit effective noise pattern capture and improved learning capabilities, especially with limited training samples~\cite{alamdari2021improving, germain2018speech, xu2020listening, Saleem2020DeepNN, kashyap2021speech}. The vision transformer (ViT)~\cite{dosovitskiy2021image}, such as CleanUNet, swin transformer, and CrossVit~\cite{kong2021speech, chen2021crossvit}, has found success in various domains, including audio adaptation tasks. The audio spectrogram transformer showed the potential of attention-based models in audio classification~\cite{gong2021ast}. Gul and Khan~\cite{gul2023survey} explored various audio enhancement algorithms, utilizing image U-Net for quality improvement across multiple domains.

In recent years, deep learning applications in bird sound denoising have been extensively explored. Li et al.,~\cite{li2023deeplabv3+} introduced the Deeplabv3+ vision transformer, emphasizing perceptual quality enhancement for noisy audio signals. Zhang and Li~\cite{zhang2023complex} proposed a complex image generation SwinTransformer network, addressing challenges in high-performance audio denoising while emphasizing the quality of generated time-frequency representations. Additionally, Zhang et al.,~\cite{zhang2024influence} explored the influence of recording devices and environmental noise on acoustic index scores, providing implications for bird sound-based assessments. These studies collectively contribute to advancing bird sound denoising through innovative deep-learning approaches~\cite{li2023bio, zhang2024automatic, li2023dcht, garciacomplex, li2023dpatd}. We assessed our model using ground truth provided by~\cite{zhang2024influence}. The dataset was sourced from natural environments, introducing a potential bias if noise bandwidths perfectly match bird sounds. Most noises have distinct shapes from bird sounds and are removed by our model.

 %

\begin{figure*}
	\centering
	\includegraphics[width=1.0\textwidth]{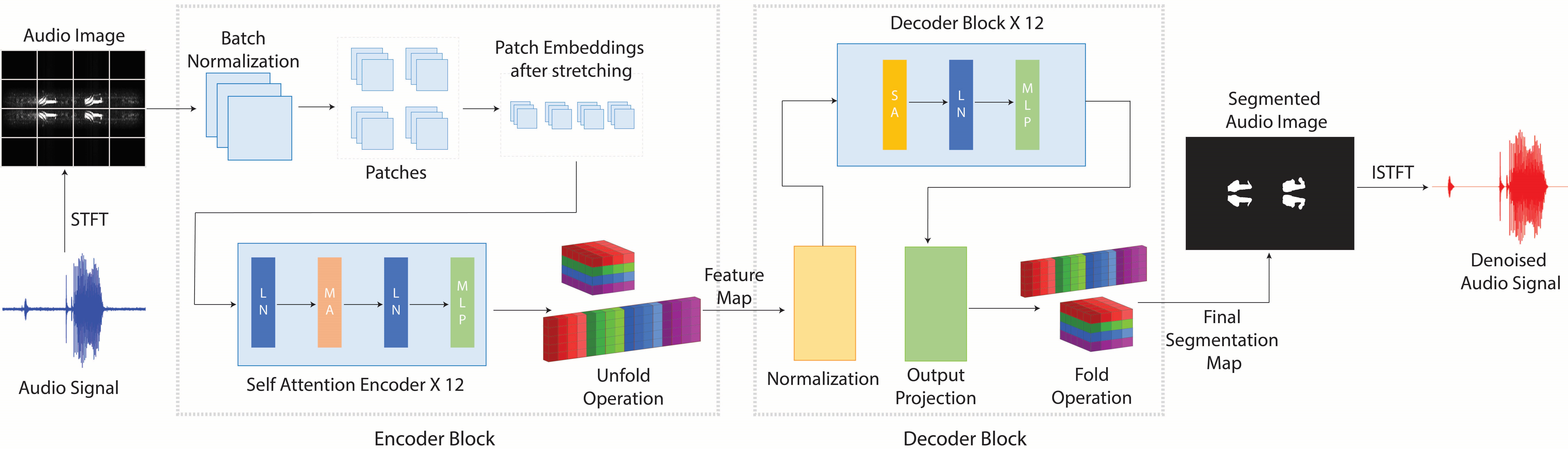}
	\caption{The overview of our ViTVS architecture. The encoder comprises a sequence of self-attention encoder blocks, each executing normalization, patch creation, and embedding layers. The decoder mirrors the encoder with additional operations, including unfolding and output projection, culminating in the final segmentation map. Both encoder and decoder consist of 12 blocks.}
 \vspace{-0.3cm}
	\label{Fig:arc}
\end{figure*}

We present a vision transformer for visual (ViTVS) bird sound denoising model, as shown in Fig.~\ref{Fig:arc}. By transforming audio denoising into an image segmentation problem, ViTVS harnesses the capabilities of the vision transformer (ViT) architecture. The model includes essential modules: Short-Time Fourier Transform (STFT) for audio-to-image conversion, a vision transformer-based ViTVS model for segmentation, and Inverse Short-Time Fourier Transform (ISTFT) for audio reconstruction.
ViTVS generates comprehensive, long-range, and multi-scale representations, distinguishing itself from conventional methods. The transformer-based encoder-decoder architecture, exemplified by PtDeepLab~\cite{li2023deeplabv3+}, effectively captures and fuses multi-scale features, showcasing its prowess in audio image segmentation. Our contributions are three-fold:
\begin{enumerate}

\item We introduce a vision transformer (ViTVS) model for bird sound denoising. Unlike existing ViT models, ViTVS redefines audio denoising as an image segmentation challenge, showcasing its distinctive ability to handle intricate bird sound patterns. The encoder-decoder architecture of ViTVS effectively captures and fuses multi-scale features, surpassing the capabilities of traditional ViT methods. 
\item ViTVS encompasses important components by adding s PatchEmbedding layer to transform input audio into image-like patches and developing a SelfAttentionEncoderBlock for multi-head attention.  

\item ViTVS achieves state-of-the-art bird sound denoising performance and intricate pattern learning. Leveraging negative log-likelihood loss enhances denoising capabilities, establishing ViTVS as a benchmark.
\end{enumerate}

\section{Methods}

\subsection{Problem}
A noisy audio signal $y(t)$ is modeled as the sum of a clean audio signal $x(t)$ and additive noise $\varepsilon(t)$~\cite{li2020speech}:
\begin{equation}
  y(t) = x(t) + \varepsilon(t)  
\end{equation}
The objective of audio denoising is to learn a mapping $\mathcal{F}$ that extracts the clean audio component $X$ from the mixture audio signal $Y = \{y_i\}_{i=1}^{N}$, minimizing the approximation error between the estimated denoised audio $\mathcal{F}(Y)$ and the clean audio $X$. Our approach also transforms audio denoising into an image segmentation task, aiming to minimize the error between predictions of audio images $\mathcal{I} = \{I_i\}_{i=1}^{N}$ from our model and their corresponding ground truth labeled masks $U = \{u_i\}_{i=1}^{N}$. 

\subsection{Motivation}

Our motivation stems from the limitations of previous bird sound denoising methods~\cite{zhang2023birdsoundsdenoising}, which struggle to effectively separate bird calls from background noise. Traditional filtering techniques and general-purpose deep learning models are not designed specifically for bird sound tasks. To address these challenges, we propose the ViTVS model, tailored to capture the unique characteristics of avian vocalizations. By transforming audio denoising into an image segmentation task using a novel vision transformer model, we aim to improve denoising performance and accuracy. Our research seeks to advance techniques for analyzing and processing bird vocalizations.

\subsection{Preliminary}
Our model utilizes Short-Time Fourier Transform (STFT) and Inverse Short-Time Fourier Transform (ISTFT) for analyzing and reconstructing audio signals~\cite{wang2021denoising}. We use the absolute value of STFT to segment and identify noise masks, while preserving the original phase in ISTFT for accurate time-domain signal reconstruction. This technique ensures effective noise removal by processing the audio into frames of short-term spectra.

\noindent \textbf{Audio Image Construction:} STFT is suitable for non-stationary audio signals since it is capable of dividing the signal into frames and analyzing their short-term spectra:
\begin{equation}\label{eq:a}
    I = \left| \text{STFT}_y(t, f) \right|.
\end{equation}
where $(I)$ represents the construction of an audio image at time $t$ and frequency $f$.
We generate audio images using STFT that can convert bird sound audio to image representations.

\noindent \textbf{Denoised Audio Reconstruction:} After noise removal in the frequency domain, denoised audio is reconstructed using ISTFT. Filtering noise areas in the frequency domain is guided by predicted masks ($\overset{\sim}{u} = F(I)$)~\cite{zhang2023birdsoundsdenoising}:
\begin{equation}\label{eq:O}
    \mathcal{O'} = \mathcal{O},\quad \text{and} \quad \mathcal{O'}[\overset{\sim}{u}<1] = 0,
\end{equation}
where $\mathcal{O}$ is the original audio image, $\mathcal{O'}$ wil get the predicted denoised masks, Eq.~\eqref{eq:ISTFT} reconstructs denoised audio ($\hat{y}(t)$) using ISTFT after frequency domain noise removal.
\begin{equation}\label{eq:ISTFT}
    \hat{y}(t) = \text{ISTFT}(\mathcal{O}')
\end{equation}


\section{Model Architecture}


\subsection{Input Transformation}

Firstly, the model undergoes a meticulous input transformation involving batch normalization and image-to-patch conversion. The input image $I$ experiences normalization through Batch Normalization (BN), as depicted in Eq.~\eqref{eq:bn}:
\begin{equation}\label{eq:bn}
\text{BN}(X) = \frac{X - \mu}{\sqrt{\sigma^2 + \epsilon}} \cdot \gamma + \beta,
\end{equation}
where $X$ is the input, $\mu$ and $\sigma$ are mean and standard deviation, $\epsilon$ is a stability constant, $\gamma$ is the scale parameter, and $\beta$ is the shift parameter.
Subsequently, $I$ undergoes linear embedding, resulting in a curated sequence $X \in \mathbb{R}^{N \times D}$, as expressed in Eq.~\eqref{eq:linear_embedding}:
\begin{equation}\label{eq:linear_embedding}
X = \text{Linear}(\text{ITP}(\text{BN}(I))),
\end{equation}
where ITP denotes ImageToPatches operation.

\noindent \textbf{Image-to-Patch Conversion:} The essential image-to-patch function transforms $I$ into non-overlapping patches of size $p \times p$, as defined in Eq.~\eqref{eq:image_to_patches}:
\begin{equation}\label{eq:image_to_patches}
\textbf{ITP}(I)_{mn} = \textbf{flatten}(I_{(pm):(pm+p-1), (pn):(pn+p-1), :}),
\end{equation}
where $I_{(pm):(pm+p-1), (pn):(pn+p-1), :}$ selects the patch at position $(m, n)$ in the image, and $\textbf{flatten}$ reshapes this patch into a flattened vector.
We ensure proper handling of patch edges during the reassembly of these patches into a complete image.

\subsection{Encoder}
In ViTVS model, the encoder and decoder play crucial roles in their architectures. The encoder serves as the backbone, consisting of self-attention-based blocks meticulously organized into 12 layers.
\noindent \textbf{Self-Attention Encoder Block:} Each encoder block ($i$) includes self-attention mechanisms and feedforward neural networks~\cite{vaswani2023attention}, complemented by layer normalization. The progression within an encoder block is expressed in Eqs.\eqref{eq:attention_output},~\eqref{eq:mlp_output}, and~\eqref{eq:x_update}:
\begin{align}
\text{Attention\_Output}_i &= \text{MultiheadAttention}(\text{LN}(X_i)) \label{eq:attention_output}, \\
\text{MLP\_Output}_i &= \text{MLP}(\text{LN}(\text{Attention\_Output}_i)) \label{eq:mlp_output}, \\
X_{i+1} &= X_i + \text{MLP\_Output}_i, \label{eq:x_update}
\end{align}

\noindent \textbf{Multihead Self-Attention:} The intricate operation of queries $Q$, keys $K$, and values $V$ unfolds through linear projections of $X$. The nuances of this process are encapsulated in Eqs.~\eqref{eq:multihead_attention} and~\eqref{eq:attention_scores}:
\begin{equation}\label{eq:multihead_attention}
\begin{aligned}
Q &= X \cdot W_Q, &\quad
K &= X \cdot W_K, &\quad
V &= X \cdot W_V,
\end{aligned}
\end{equation}
\begin{equation}\label{eq:attention_scores}
\text{Attention\_Scores} = \text{softmax}\left(\frac{QK^T}{\sqrt{D_k}}\right) \cdot V,
\end{equation}
where $X$ is the input sequence at a certain layer, $W_Q$, $W_K$, and $W_V$ are learnable weight matrices and $D_k$ is the dimension of queries and keys~\cite{vaswani2023attention}.

\noindent \textbf{Feedforward Neural Network (MLP):} The Multi-Layer Perceptron (MLP) works with two linear layers and GELU activation, as portrayed in Eq.~\eqref{eq:mlp}:
\begin{equation}\label{eq:mlp}
\text{MLP}(X) = \text{GELU}(X \cdot W_1 + b_1) \cdot W_2 + b_2,
\end{equation}
where  $W_{1}$, $W_{2}$, $b_{1}$, and $b_{2}$ represent the weights matrix and biases vectors that the model learns through training to make predictions based on the input $X$. The subscripts $1$ and $2$ denote the layers of the MLP.

\subsection{Decoder}

The decoder mirrors the encoder structure but incorporates an output projection layer, also featuring 12 layers.

\noindent \textbf{Decoder Block:} Each decoder block entails a self-attention encoder block, layer normalization, and an MLP. The baton passes from one block to the next, as expressed in Eqs.~\eqref{eq:decoder_attention_output}, ~\eqref{eq:decoder_mlp_output}, and~\eqref{eq:decoder_x_update}:
\begin{align}
\text{Attention\_Output}_i &= \text{SelfAttentionEncoderBlock}(\text{LN}(X_i)) \label{eq:decoder_attention_output}, \\
\text{MLP\_Output}_i &= \text{MLP}(\text{LN}(\text{Attention\_Output}_i)) \label{eq:decoder_mlp_output}, \\
X_{i+1} &= X_i + \text{MLP\_Output}_i ,\label{eq:decoder_x_update}
\end{align}

\noindent \textbf{Output Projection:} The final step involves projecting the output to the original image dimensions through a linear layer, followed by a folding process, as defined in Eq.~\eqref{eq:output_projection}:
\begin{equation}\label{eq:output_projection}
\text{Output} = \text{Fold}(\text{Linear}(X)).
\end{equation}

In short, both the encoder and decoder architectures comprise 12 layers, each housing self-attention blocks, layer normalization, and multi-layer perceptrons (MLPs).

\subsection{Loss Function and Algorithm}

\noindent \textbf{Loss Function:} We train our ViTVS model using a negative log-likelihood loss (NLL)~\cite{zhu2018negative}, expressed in Eq.~\eqref{eq:loss}:
\begin{equation}\label{eq:loss}
\text{Loss} = -\frac{1}{N} \sum_{i=1}^{N} \log\left(\frac{\exp(\text{Output}_i[M_i])}{\sum_{j=1}^{C} \exp(\text{Output}_i[j])}\right),
\end{equation}
where $N$ denotes the number of pixels, $C$ signifies the number of classes, $j$ is a summation index representing each class ($1$ to $C$). and $M_i$ represents the ground truth label for pixel $i$.

This loss function quantifies dissimilarity between predicted and ground truth labels, fostering the model to generate accurate and contextually relevant segmentation maps. The negative log-likelihood~\cite{zhu2018negative} penalizes incorrect predictions, contributing to a robust learning process. The log-softmax ensures numerical stability during computation.


\noindent \textbf{Training Algorithm.}
We train our ViTVS model using Alg.~\ref{alg:VitSegmentation}. 
\begin{algorithm}[h]
   \caption{Vision Transformer Segmentation for Visual Bird Sound Denoising}
   \label{alg:VitSegmentation}
   \begin{algorithmic}[1]
      \REQUIRE Noisy audio signals $Y=\{y_i\}_{i=1}^{N}$ and labeled mask images $M=\{m_i\}_{i=1}^{N}$, where $N$ is the total number of audio samples.
      \ENSURE Denoised audio signals $\mathcal{F}(Y)$
      \STATE Generate audio images $\mathcal{I}$ using Eq.~\eqref{eq:a}
      \STATE Initialize vision transformer model with encoder (12 blocks), decoder (12 blocks), and likelihood loss function.
      \FOR{$\text{iter} = 1$ \TO $II$}
         \FOR{$k = 1$ \TO $n_B$}
            \STATE Extract batch-wise data: $I^k$ and $M^k$ sampled from $\mathcal{I}$ and $M$ using Eqs.~\eqref{eq:bn}, ~\eqref{eq:linear_embedding} and ~\eqref{eq:image_to_patches}.
            \STATE Encode audio images using the Vision Transformer encoder with 12 blocks using Eqs.~\eqref{eq:attention_output}, ~\eqref{eq:mlp_output}, and  ~\eqref{eq:x_update}.
            \STATE Obtain segmentation using the vision transformer decoder with 12 blocks using Eq.~\eqref{eq:output_projection}.
            \STATE Calculate likelihood loss using the specified likelihood loss function using Eq.~\eqref{eq:loss}.
            \STATE Backward pass and update model parameters.
         \ENDFOR
      \ENDFOR
      \STATE Obtain the clean frequency domain using Eq.~\eqref{eq:O}
      \STATE Output the denoised audio signals using Eq.~\eqref{eq:ISTFT}
   \end{algorithmic}
\end{algorithm}



\section{Experiments}

\subsection{Implementation details}

To ensure the efficacy and robustness of our approach, we utilized PyTorch version 2.0.1 in conjunction with torchvision version 0.15.2. This framework was complemented by CUDA version 11.7, leveraging parallel processing capabilities on the NVIDIA GeForce RTX 4070 GPU. 
The learning rate is set to $5e-5$ with an AdamW optimizer, and the weight decay is adjusted to 5e-4. Further, we resize input images to 256 × 256 × 3 with a mini-batch size of 8 for 100 epochs. 

\begin{figure*}
	\centering
	\includegraphics[width=0.9\textwidth]{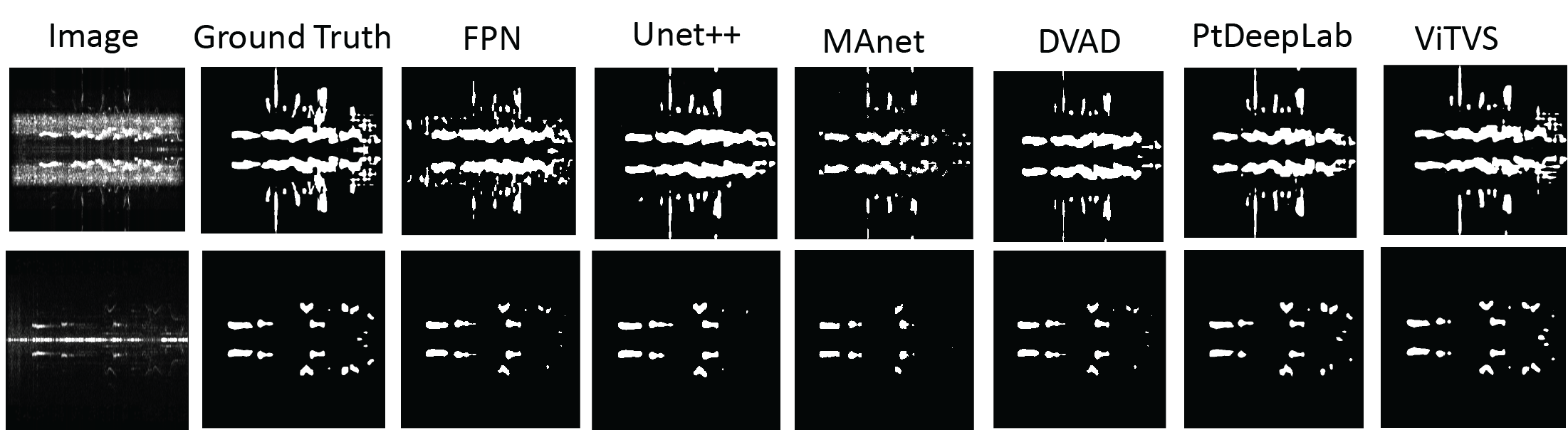}
	\caption{Segmentation results comparisons. The leftmost column is the original audio image. The ground truth is the labeled mask. }\label{Fig:mask}
\vspace{-0.3cm}
\end{figure*}

\subsection{Datasets}

Our model is evaluated using the BirdSoundsDenoising dataset, which contains 14,120 audio signals lasting one to fifteen seconds each, split into training (10,000), validation (1,400), and testing (2,720) sets~\cite{zhang2023birdsoundsdenoising}. This dataset stands out by featuring natural environmental noises—like wind, waterfalls, and rain—instead of artificial noise.


\subsection{Results}


We evaluate our ViTVS model against existing methods in terms of learning efficacy, capability, and qualitative outcomes. We use three key metrics—F1, IoU, and Dice—for image segmentation performance assessment~\cite{zhang2023birdsoundsdenoising}, and the signal-to-distortion ratio (SDR) for audio denoising efficacy, as outlined in~\cite{li2023deeplabv3+}. Higher scores in these metrics indicate better segmentation and denoising performance, reflecting closer alignment with ground truth, which calculated based on labeled masks rather than real-world clean bird sounds.  As shown in Tab.~\ref{tabcom}, Fig.~\ref{Fig:arc}, and Fig.~\ref{Fig:mask}, our model significantly enhances denoising performance, paving the way for applications in bird species classification and audio signal enhancement.

\subsection{Performance Comparison}

Fig.~\ref{Fig:mask} shows the comparisons of five best baseline segmentation models. The performance of various audio denoising methods is detailed in Tab.~\ref{tabcom}, including $F1$ score, Intersection over Union ($IoU$), Dice score, and $SDR$ on both validation and test sets. Among these methods, ViTVS stands out, demonstrating robust denoising capabilities. ViTVS achieves the highest $F1$, $IoU$ and $Dice$ scores on both validation and test sets, 
highlighting its superior segmentation accuracy. The $SDR$ values reinforce ViTVS's excellence in denoising, positioning it as a compelling choice for audio denoising. ViTVS consistently outperforms state-of-the-art methods across multiple metrics, emphasizing its efficacy and potential for real-world applications.


To investigate architectural variations and determine the optimal configuration, we conducted an ablation study on our Vision Transformer Segmentation (ViTVS) model. The study focused on comparing variants, specifically the number of blocks in both the encoder and decoder. We trained five variants with block configurations ranging from 4 to 20, evaluating their performance on validation and test sets using $IoU$, $Dice$, and $F1$ scores. Notably, increasing blocks from 4 to 12 substantially improved denoising capabilities, emphasizing the pivotal role of block count. Intriguingly, further experimentation with 16 blocks did not yield additional benefits, suggesting 12 blocks strike an optimal balance between complexity and performance. The ViTVS 12-block variant showcased superior results compared to its 4, 5, 9, 16, and 20-block counterparts in terms of IoU, Dice, and F1 scores, underscoring the significance of a finely tuned model for optimal denoising. These compelling findings, detailed in Tab.~\ref{tab:ablation}, highlight the critical importance of selecting the proper number of blocks for better results.

\begin{table}[t]
\small
\begin{center}
\captionsetup{font=small}
\caption{Results comparisons of different methods ($F1, IoU$, and $Dice$ scores are multiplied by 100.}
\label{tabcom}
\vspace{-0.2cm}
\setlength{\tabcolsep}{+0.3mm}{
\begin{tabular}{lllll|lllllllll}
\hline \label{tab:md}
 \multirow{2}{*}{Networks}
 &  \multicolumn{4}{c}{Validation} & \multicolumn{4}{c}{Test} \\
 \cmidrule{2-9}
& $F1$ & $IoU$ & $Dice$ & $SDR$ & $F1$ & $IoU$ & $Dice$ & $SDR$ \\
\hline
U$^2$-Net~\cite{qin2020u2}  &60.8 &45.2 &60.6 &7.85 & 60.2  &44.8 &59.9 & 7.70\\
MTU-NeT~\cite{wang2022mixed}  &69.1 &56.5 &69.0  &8.17 & 68.3  &55.7 & 68.3 &7.96  \\
Segmenter~\cite{strudel2021segmenter} & 72.6  & 59.6 & 72.5 & 9.24 & 70.8 & 57.7 & 70.7 & 8.52   \\
U-Net~\cite{ronneberger2015u}  &75.7 &64.3 &75.7 & 9.44 &74.4 &62.9 &74.4 & 8.92    \\
SegNet~\cite{badrinarayanan2017segnet}  &77.5 &66.9 &77.5 & 9.55&76.1 &65.3 &76.2 & 9.43 \\
YOLOv8~\cite{wang2022mixed} & 74.5 & 62.1 & 74.5 & 8.33 & 73.6 & 60.9 & 73.3 & 8.12 \\
FPN~\cite{lin2017feature} & 82.4 & 70.1 & 82.5 & 9.91 & 81.2 & 69.3 & 81.2 & 9.64 \\
Unet++~\cite{zhou2018unet} & 83.0 & 71.0 & 83.1 & 9.98 & 82.1 & 69.8 & 82.1 & 9.87 \\
MAnet~\cite{Li_2022} & 83.5 & 71.7 & 83.5 & 10.12 & 82.4 & 70.9 & 82.4 & 9.92 \\
DVAD~\cite{zhang2023birdsoundsdenoising} & 82.6 & 73.5 & 82.6 & 10.33 & 81.6 & 72.3 & 81.6 & 9.96 \\
PtDeepLab~\cite{li2023deeplabv3+} & 83.4 & 75.9 & 83.4 & 10.49 & 83.1 & 75.4 & 83.0 & 10.43\\
\hline
\textbf{ViTVS} & \textbf{88.3} & \textbf{80.9} & \textbf{88.3} & \textbf{11.04} & \textbf{90.7} & \textbf{80.0} & \textbf{87.6} & \textbf{10.86} \\
\hline
\end{tabular}}
\end{center}
\vspace{-.3cm}
\end{table}


\begin{table}[ht]
\small
\begin{center}
\caption{Ablation study results comparing ViTVS variants.}
\label{tab:ablation}
\vspace{-0.2cm}
\setlength{\tabcolsep}{+0.3mm}{
\begin{tabular}{l|lll|lll}
\hline 
\multirow{2}{*}{Model Variant}
& \multicolumn{3}{c}{Validation} & \multicolumn{3}{c}{Test} \\
\cmidrule{2-7}
& $IoU$ & $Dice$ & $F1$ & $IoU$ & $Dice$ & $F1$ \\
\hline
ViTVS 4-block & 60.3 & 72.6 & 72.7 & 54.5 & 68.4 & 68.4 \\
ViTVS 5-block & 68.8 & 76.7 & 76.7 & 67.9 & 76.0 & 76.0 \\
ViTVS 9-block & 76.8 & 84.7 & 84.7 & 75.2 & 84.0 & 84.0 \\
\textbf{ViTVS 12-block}  & \textbf{80.9} & \textbf{88.3} & \textbf{88.3} & \textbf{80.0} & \textbf{87.6} & \textbf{90.7} \\
ViTVS 16-block & 78.5 & 86.2 & 86.3 & 78.1 & 85.5 & 85.5 \\
ViTVS 20-block & 80.6 & 88.1 & 76.7 & 79.9 & 87.5 & 87.5 \\
\hline
\end{tabular}}
\end{center}
\vspace{-.6cm}
\end{table}



\section{Discussion}
Our ViTVS model demonstrates superior performance compared to the CNN-based PtDeepLab model~\cite{li2023deeplabv3+} and eleven other state-of-the-art segmentation and denoising methods. It excels in segmentation results and enhances representation capacity by effectively capturing multi-scale information for detailed context patterning. The ablation study results, shown in Tab.~\ref{tab:ablation}, and the segmented mask quality in Fig.~\ref{Fig:mask} validate our model's robustness in preserving essential clean signals. While converting audio denoising to segmentation proposed in~\cite{zhang2023birdsoundsdenoising}, our unique bird vision-transformed segmentation model marks the first of its kind to notably enhance denoising tasks.

\section{Conclusion}
In this paper, we introduce a vision transformer for a visual bird sound denoising model (ViTVS). Our ViTVS model is designed with a self-attention mechanism, patch-based processing, layer normalization, and multi-layer perceptrons to capture long-range dependencies and effectively handle diverse input sizes. ViTVS exhibits superior performance to state-of-the-art methods, as evidenced by extensive experimental evaluations, including comparisons with CNN-based self-attention and diffusion models. Therefore, our developed ViTVS model is suitable for bird sound denoising. 



\bibliographystyle{IEEEtran}
\bibliography{denoised_2}

\end{document}